\documentclass[twocolumn,a4paper,superscriptaddress,floatfix,showpacs]{revtex4}
\usepackage{amsmath,amssymb,epsfig,color,textcase}

\begin{document}

\title{Theoretical prediction of Jahn-Teller distortions and orbital ordering in 
Cs$_2$CuCl$_2$Br$_2$}
\author{S.V.~Streltsov}
\affiliation{Institute of Metal Physics, S.Kovalevskoy St. 18, 620990 Ekaterinburg, Russia}
\affiliation{Ural Federal University, Mira St. 19, 620002 Ekaterinburg, Russia}
\email{streltsov@imp.uran.ru}

\author{D.I.~Khomskii}
\affiliation{II. Physikalisches Institut, Universit$\ddot a$t zu K$\ddot o$ln,
Z$\ddot u$lpicher Stra$\ss$e 77, D-50937 K$\ddot o$ln, Germany}


\date{\today}

\begin{abstract}
With the use of the density function calculations we show that the actual crystal
structure of Cs$_2$CuCl$_2$Br$_2$ should contain elongated in the $ab-$plane
CuCl$_4$Br$_2$ octahedra, in contrast to the experimentally observed compression
in $c-$direction. We also predict that the spins on Cu$^{2+}$ ions should
be ferromagnetically ordered in $ab-$plane, while the exchange interaction
along $c-$direction is small and its sign is uncertain.
\end{abstract}

\maketitle

\section{Introduction \label{intro}}
The system Cs$_2$Cui(Cl,Br)$_4$ attracts a lot of attention
mainly because of its unconventional magnetic properties. 
Cs$_2$CuBr$_4$ is the only triangular-lattice antiferromagnet, which 
exhibits unusual quantum magnetization plateau,~\cite{Fortune2009}
while Cs$_2$CuCl$_4$ shows field-induced Bose-Einstein condensation of 
magnons.~\cite{Radu2005} Thus the investigation of intermediate 
compositions of Cs$_2$CuCl$_{4-x}$Br$_x$ may not only shed some light
on the magnetic properties of both compounds, but may also results
in the discovery of new phenomena.

The mixed Cs$_2$CuCl$_{4-x}$Br$_x$ crystal series was successfully
grown at 50 $^{\circ}$C with orthorombic crystal structure.  
However, at room temperature in the synthesis process the stabilization of a 
new tetragonal phase for $1<x<2$ was observed.~\cite{Kruger2010}

\begin{figure}[b!]
 \centering
 \includegraphics[clip=false,width=0.4\textwidth]{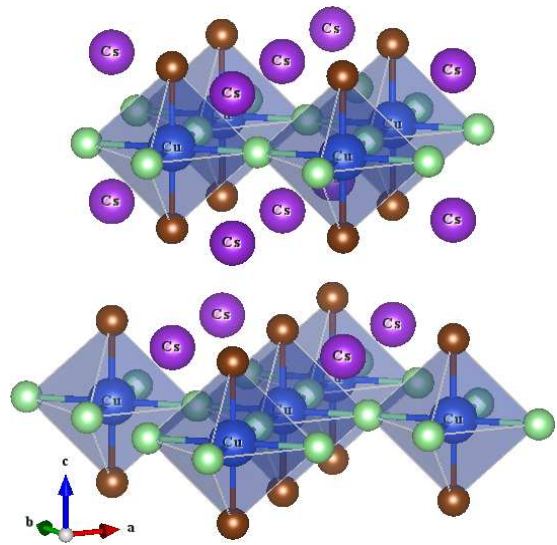}
\caption{\label{cryst.str}(color online). The crystal structure of
the tetragonal phase of Cs$_2$CuCl$_2$Br$_2$. Blue balls are
Cu ions, light green, brown and purple are Cl, Br and Cs
ions respectively. According to Ref.~\onlinecite{Kruger2010},  the 
CuCl$_4$Br$_2$ octahedra are compressed
along the c--axis. The image was generated using VESTA
software.~\cite{MommaK.Izumi2011}}
\end{figure}

The crystal structure of tetragonal specimens consists
of the CuCl$_2$ layers stacking in the $c-$direction 
and divided by the Cs and Br atoms, see Fig.~\ref{cryst.str}.
It is important to mention that Cu ions in one of the
layers placed on a top/bottom of the void between CuCl$_4$ plaquettes
of another layer. According ro Ref.~\onlinecite{Kruger2010}, the Cl and 
Br ions form octahedra surrounding Cu with
two short, apical, Cu-Br and four long, planar, Cu-Cl 
bonds. Such kind of the compressed 
octahedra are quite untypical for the Jahn-Teller
Cu$^{2+}$ ions with $d^9$ electronic configuration 
for the dense crystal structures,
and the authors of Ref.~\onlinecite{Kruger2010} mentioned that the real type
of distortions may be hidden by the multidomain structures.  

There are a lot of insulating materials, where ligand octahedra
surrounding Jahn-Teller active metal ion turn out
to be elongated: KCrF$_4$~\cite{Margadonna2006,Giovannetti2008},
Cs$_2$AgF$_4$~\cite{McLain2006,Wu2007}, K$_2$CuF$_4$\cite{Ito1976,KK1973}, 
and others, while there are only few systems with the 
opposite distortion. Generally speaking there can be 
different mechanisms, which stabilize elongated octahedra.~\cite{Kanamori1960}
One of the reasons can be the gain in the magnetic energy
due to more efficient hoppings between half-filled
$x^2-y^2$-like and oxygen $p-$orbitals in the case of the $d^9$
configuration. 
Another one is related with the features of the elastic
interactions. It can be shown that the total 
energy of elongated octahedra is lower than compressed
if in the expression for the elastic energy the terms 
of higher order than quadratic are taken into 
account.~\cite{KhomskiiBrink2000}

In the present paper, using ab-initio band structure
calculations, we found that the crystal structure of
Cs$_2$CuCl$_2$Br$_2$ ($x=2$), which corresponds to the lowest total energy,
indeed corresponds to the elongated CuCl$_4$Br$_2$
octahedra. This is in contrast to the observed 
experimental structure,~\cite{Kruger2010} but support
general tendency in the Jahn--Teller distortions to
stabilize elongated, not compressed octahedra.
The more detailed structural study should be carried out
to confirm (or disprove) the predicted lattice and orbital 
ordering.

\section{Calculation details}
The pseudo-potential PWscf code was chosen for the 
calculations.~\cite{Giannozzi2009}  We used ultrasoft 
pseudo-potentials with the nonlinear core correction and the 
Perdew-Burke-Ernzerhof (PBE) version of
the exchange-correlation potential.~\cite{Perdew1996}
In order to take into account strong Coulomb repulsion
on the Cu sites the GGA+U approximation was utilized.~\cite{Anisimov1997}
On-site Coulomb repulsion parameter U was chosen to be 7.0~eV, while
intra-atomic Hund's rule exchange $J_H=0.9$~eV.~\cite{Liechtenstein1995}

The charge density and kinetic energy cut-offs equal 40 Ry and
200 Ry, respectively. 144 $k-$points (6$\times$6$\times$4) in a full 
part of the Brillouin zone for the unit cell, consisting of 4 
formula units (f.u.), were used in the self-consistency course. 
The structural optimization was performed while each component of the force 
were more than 2 mRy/a.u. No symmetry operations were used in the course 
of self-consistency.

The crystal structure was taken for Cs$_2$CuCl$_{2.2}$Br$_{1.8}$.\cite{Kruger2010}
We used the supercell $\sqrt2 \times \sqrt 2 \times 2$ cell to allow the 
simplest types of the orbital ordering.

\section{Calculation results}

The total energy is known to depend on the
type of magnetic ordering, and this magnetic ordering may
have an influence on the lattice distortions through
the stabilization of the particular orbital ordering.~\cite{KK-UFN} 
That is why it is important to study possible lattice distortions 
together with the analysis of the magnetic interactions.

The total energies of the following magnetic configuration
were calculated: nonmagnetic (NM), ferromagnetic (FM), 
AFM-A type (when all ions in $ab-$plane are
ferromagnetically ordered, while the interaction between
planes is antiferromagnetic), and AFM-C type
(nearest Cu in the $ab-$plane are AFM ordered, 
the next nearest neighbors in $c-$direction are ferromagnetically 
coupled).

\begin{table}
\centering \caption{\label{TotalEnergies}
Total energies in meV per formula unit for different magnetic 
configurations for the experimental crystal structure.~\cite{Kruger2010}
The energy of the FM configuration was chosen as zero.}
\vspace{0.2cm}
\begin{tabular}{l|ccccc}
\hline
\hline
               & Total energy, meV  \\
\hline
FM    & 0\\
AFM-A & 3.7\\
AFM-C & 2.9\\
NM    & 8.8\\
\hline
\hline
\end{tabular}
\end{table}

The lowest in energy in the tetragonal crystal structure of 
Ref.~\onlinecite{Kruger2010} 
turns out to be the FM configuration (see Tab.~\ref{TotalEnergies}).
The analysis of the occupation matrix shows that the single
hole in $3d-$shell of Cu$^{2+}$ is localized on 
$3z^2-r^2-$orbital in the FM configuration. Such
an orbital filling is obviously a result of the local
compression of CuCl$_4$Br$_2$ octahedra in the structure of Ref.~\onlinecite{Kruger2010}. 

The ground state is metallic for 
any of the investigated magnetic configurations.
This is in contrast to the fact that the undoped materials Cs$_2$CuCl$_4$ 
and Cs$_2$CuBr$_4$ are Mott insulators~\cite{Vachon2008,Foyevtsova2009},
and the samples of the intermediate compound Cs$_2$CuCl$_{2.4}$Br$_{1.6}$
do not have metallic shine.~\cite{Kruger2010}

\begin{figure}[t!]
 \centering
 \includegraphics[clip=false,width=0.5\textwidth]{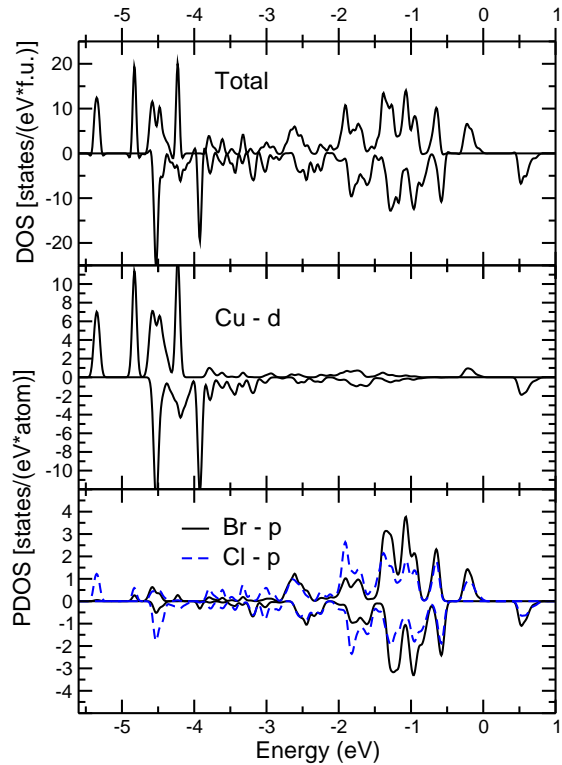}
\caption{\label{DOS}(color online). The total and partial
DOS for FM configuration in the relaxed crystal structure
with alternating long and short Cu-Cl bond in the $ab-$plane. 
Positive (negative) values correspond to spin majority (minority) 
states. The Fermi level is set to zero.}
\end{figure}

The compression of the octahedra in Cs$_2$CuCl$_2$Br$_2$ in the $c-$direction 
prevents the $x^2-z^2$/$y^2-z^2$  orbital order observed in 
KCuF$_3$~\cite{Anisimov1997} and K$_2$CuF$_4$~\cite{Ito1976,KK1973}, where Cu ions 
also have one hole in the $e_g$ sub-shell.
In order to allow the same type of the orbital pattern and compare total energies
of different solutions we tetragonally distorted CuCl$_4$Br$_2$ in the $ab-$plane
and relaxed crystal structure with the constrain to keep the
same cell volume. In addition we also performed the lattice 
optimization for the initial non-distorted in $ab-$plane
structure, since the experimental structure not
necessarily corresponds to the ground state crystal
structure in the chosen approximation (GGA+U). In effect
we obtained two crystal structures corresponding to the 
same magnetic order (relaxed initial experimental and
relaxed distorted in ab-plane structures). The results are summarized in
Tab.~\ref{TotalEnergiesR}, where the second, forth, and fifth lines
refer to structure optimized staring from the lattice presented
in Ref.~\onlinecite{Kruger2010} (four equal Cu-Cl distances)
and the first and the third lines  correpond to the results
for the structure with octahedra elongated in ab-plane.

\begin{figure}[t!]
 \centering
 \includegraphics[clip=false,width=0.5\textwidth]{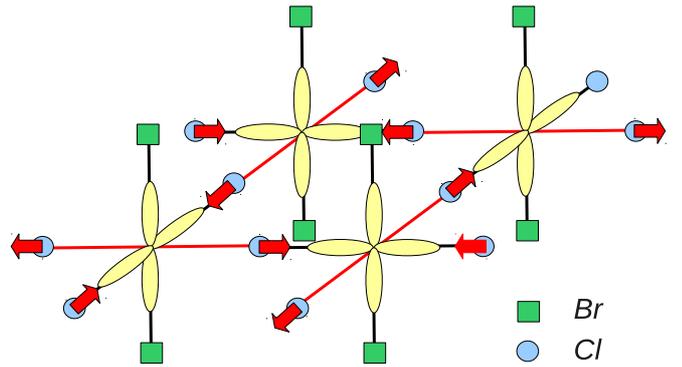}
\caption{\label{OO}(color online). The sketch of the proposed orthorombic
distortions corresponding to the lowest total energy in Tab.~\ref{TotalEnergiesR}, 
when all CuCl$_4$Br$_2$ octahedra are elongated in alternating directions
in the $ab$ plane (long Cu-Cl bonds
are shown in red) and the orbital order, which is expected for such kind
of the distortions (hole orbitals of Cu$^{2+}$ are shown).
}
\end{figure}

The lowest total energy corresponds to
FM and AFM-A type configurations with strongly distorted in the 
$ab-$plane CuCl$_4$ plaquettes, as shown in Fig.~\ref{OO}. The energy difference between these
two solutions is tiny and may depend on the details of the calculations,
but both lie much lower ($\sim$ 558 meV) in energy than the ground state
magnetic configuration (FM) for the tetragonal not optimized crystal structure.
Moreover, it is clearly seen from Tab.~\ref{TotalEnergiesR} that only half of 
this difference can be compensated by the ionic relaxation which
does not change a local symmetry of Cu$^{2+}$ ions (i.e. that there
are 4 equal Cu-Cl bonds in $ab-$plane).

The second half of the total energy decrease is related to the
distortions in $ab-$plane such that they do not change
the average Cu-Cl bond length, but create checkerboard order
of the long and short Cu-Cl bonds. The distortions in the $ab-$plane
are accompanied by a moderate elongation of the CuCl$_4$Br$_2$
octahedra in $c-$direction, Cu-Br bond length increases on $\sim$ 0.1 \AA,
which is compensated by the Cs-Br bond squeezing.
Thus, instead of compressed in $c-$direction octahedra, the 
ionic relaxation rather stabilizes the elongated in alternating
directions in $ab-$plane CuCl$_4$Br$_2$ octahedra.

\begin{table}
\centering \caption{\label{TotalEnergiesR}
Distances and total energies per formula unit for 
different magnetic configurations for the relaxed crystal structures.
The energy of the FM configuration with two long
and two short Cu-Cl bonds was chosen as zero, it is  
557.7 meV lower than the FM solution for the
tetragonal not relaxed crystal structure.}
\vspace{0.2cm}
\begin{tabular}{l|c|c|ccc}
\hline
\hline
      & Cu-Cl dist., \AA            & Cu-Br dist., \AA  & Total energy, meV  \\ 
\hline
FM    & 3.02$\times$2/2.25$\times$2 & 2.55 $\times$2    &0    \\   
FM    & 2.64$\times$4               & 2.45 $\times$2    &285.1\\  
AFM-A & 3.02$\times$2/2.25$\times$2 & 2.55 $\times$2    &-0.8\\
AFM-C & 2.64$\times$4               & 2.44 $\times$2    &280.6\\
NM    & 2.64$\times$4               & 2.44 $\times$2    &286.4\\
\hline
\hline
\end{tabular}
\end{table}

Such type of the distortions results 
in the $x^2-z^2$/$y^2-z^2$ orbital order, like in KCuF$_3$, see
Fig.~\ref{OO}. 
This orbital pattern leads to a strong ferromagnetic 
super-exchange in $ab-$plane, which stabilizes FM or AFM-A
magnetic configurations, which agrees with our total energy
calculations.

Finally we present the total and partial density of states (DOS) plots
for the distorted in $ab-$plane relaxed crystal structure, e.g. for 
FM configuration [Fig.~\ref{DOS})]. Due to a large on-site Coulomb
repulsion Cu$-3d$ band goes away from the Fermi level and is
placed mainly in the energy range from -5.5 eV to -3 eV. The Br$-p$ and Cl$-p$ states
are concentrated from -3 eV to 1 eV. It is important to note, that in contrast
to a naive expectation, the largest contribution to the bottom of the 
conduction band (as well as to the top of the valence band) comes from
Cl$-p$ and Br$-p$ states: $\sim$3.6 states/[eV*f.u.], while
Cu$-d$ provides only 1.9 states/[eV*f.u.] (the rest belongs to other 
states of Cu, Cl and Br ions). This means that the hole is actually
localized not on the Cu-$x^2-y^2$ like orbital, but rather
on the Wannier orbital, which is centered on the Cu site, has 
$x^2-y^2$ symmetry, but also has significant contributions (tails) on 
the surrounding Cl and Br ions. This is similar to the situation
in Cs$_2$Au$_2$Cl$_6$,\cite{Ushakov2011} but here the spatial orientation of the Wannier
orbital is different. The strong admixture of Cl$-p$ and Br$-p$ states
to the conduction band may be the reason of the small band gap value 
$\sim$0.5 eV in the present GGA+U calculations, since these states act as a 
ballast. We expect that the LDA+U$_{WF}$ approximation, where U is applied 
not only on the $d-$part, but on a whole Wannier function will result in a 
larger band gap.

Summarizing, on the basis of the ab-initio calculations we have shown
that the layered material Cs$_2$CuCl$_2$Br$_2$, containing Jahn-Teller ion 
Cu$^{2+}$, which was considered earlier as a rare example of the Jahn-Teller system
with localized electrons and $e_g$ degeneracy with locally
compressed ligand octahedra, must in fact have elongated octahedra 
with the long axes alternating in the basal plane.
Thus, yet  one more Jahn-Teller material turns out to be not an exception, 
but rather follows the general rule that the octahedra around such ions
are elongated. This form of the Jahn-Teller distortions and orbital ordering
should  lead to strong ferromagnetic exchange in ab-plane, the interlayer
exchange being very weak. The predicted crystal and magnetic structure should 
be observable by the detailed structural and magnetic studies.

\section{Acknowledgments}
We thank R. Coldea who drew our attention to this system.
This work is supported by the Russian Foundation for Basic Research 
via  RFFI-10-02-96011 and RFFI-10-02-00140, by the Ural branch of Russian
Academy of Science through the young-scientist program and by
the German projects SFB 608, DFG GR 1484/2-1, FOR 1346.

\bibliography{../library}
\end{document}